\documentclass[aps,showpacs,preprintnumbers,amsmath, amssymb]{revtex4}

\usepackage{float}
\usepackage{graphics,epsfig}
\usepackage{graphicx}
\usepackage{dcolumn}
\usepackage{bm}

\begin{document}
\baselineskip=0.8 cm
\title{{\bf General holographic superconductor models with Gauss-Bonnet corrections}}

\author{Qiyuan Pan$^{1,2,3}$, Bin Wang$^{1}$}
\affiliation{$^{1}$ Department of Physics, Shanghai Jiao Tong University, Shanghai 200240, China} \affiliation{$^{2}$ Department of Physics,
Fudan University, Shanghai 200433, China} \affiliation{$^{3}$ Institute of Physics and Department of Physics, Hunan Normal University, Changsha,
Hunan 410081, China}

\vspace*{0.2cm}
\begin{abstract}
\baselineskip=0.6 cm
\begin{center}
{\bf Abstract}
\end{center}

We study general models for holographic superconductors in
Einstein-Gauss-Bonnet gravity. We find that different values of
Gauss-Bonnet correction term and model parameters can determine the
order of phase transitions and critical exponents of second-order
phase transitions. Moreover we find that the size and strength of
the conductivity coherence peak can be controlled. The ratios
$\omega_g/Tc$ for various model parameters have also been examined.

\end{abstract}

\maketitle
\newpage
\vspace*{0.2cm}

\section{Introduction}

The anti-de Sitter/conformal field theories (AdS/CFT) correspondence
\cite{Maldacena,Witten} has provided a theoretical framework to
describe the strongly coupled conformal field theories through a
weakly coupled dual gravitational description. Recently stimulated
from this correspondence, a remarkable connection between the
condensed matter and the gravitational physics has been discovered,
for reviews see \cite{HartnollRev,HerzogRev,HorowitzRev}. It was
suggested that the spontaneous $U(1)$ symmetry breaking by bulk
black holes can be used to construct gravitational duals of the
transition from normal state to superconducting state in the
boundary theory \cite{GubserPRD78}. One can look at a
(2+1)-dimensional superconductor and see its striking image
involving a charged black hole with non-trivial ``hair" in
(3+1)-dimensions \cite{HartnollPRL101}. This investigation was
further carried out beyond the probe limit by considering the
back-reaction of the field on the spacetime \cite{HartnollJHEP12}.
Gravity models with the property of holographic superconductor have
attracted considerable interest for their potential applications to
the condensed matter physics, see for example
\cite{HorowitzPRD78,Nakano-Wen,Amado,Koutsoumbas,Umeh,Maeda79,Zeng80,
Sonner,Gubser-C-S-T,Gauntlett-Sonner,Gregory,Pan-Wang,Ge-Wang,Cai-Zhang,Jing-Wang-Chen,Chen-Jing,Konoplya,Nishioka,Siopsis}.
At the moment when the condensation occurs in the boundary CFT and
in the gravitational counterpart a non-trivial hair for the black
hole is triggered, there appears a phase transition. It was argued
that this phase transition belongs to the second order
\cite{HartnollPRL101}. The phenomenological signature of this phase
transition was recently disclosed in the perturbation around such
AdS black holes \cite{bin1,bin2}.

Franco \emph{et al.} recently introduced a generalization of the
basic holographic superconductor model in which the spontaneous
breaking of a global U(1) symmetry occurs via the St\"{u}ckelberg
mechanism \cite{FrancoJHEP}. They found that a generalized
St\"{u}ckelberg mechanism of symmetry breaking allows for a
description of a wider class of phase transitions. This framework
allows tuning the order of the phase transition which can
accommodate the first order phase transition to occur, and for the
second order phase transition it allows tuning the values of
critical exponents \cite{Franco}. An interesting extension was done
in \cite{Aprile-Russo} by constructing general models for
holographic superconductivity. It was found that except some
universal model independent features, some important aspects of the
quantum critical behavior strongly depend on the choice of
couplings, such as the order of the phase transition and critical
exponents of second-order phase transitions. In addition to the
numerical investigation, analytical understanding on the phase
transition of holographic superconductor was also provided in
\cite{Herzog-2010}.

It is of great interest to generalize the investigation on the phase
transition in the holographic superconductor to the
Einstein-Gauss-Bonnet gravity. It was observed in recent works
\cite{Gregory,Pan-Wang,Ge-Wang,Brihaye} that the Gauss-Bonnet
coupling affects the condensation and the higher curvature
correction makes condensation harder to form. Further the high
curvature correction also causes the behavior of the claimed
universal ratio $\omega/T_c\approx8$ unstable. In this work we are
going to examine the effect of the Gauss-Bonnet correction on the
order of the phase transition and critical exponents. We will also
study the conductivity in a certain range of parameters and disclose
the influence given by the Gauss-Bonnet coupling.

The plan of the work is the following. In Sec. II, we will study the
phase transitions of a general class of the holographic
superconductor models via the St\"{u}ckelberg mechanism in the
Gauss-Bonnet AdS black hole background. In Sec. III we will explore
the effects of the Gauss-Bonnet correction terms and other model
parameters on the conductivity. We will conclude in the last section
of our main results.

\section{General superconducting models in Gauss-Bonnet AdS black hole}

We will consider the background solution of a neutral black hole in
$d$ dimensional Einstein-Gauss-Bonnet gravity
\cite{Boulware-Deser,Cai-2002}
\begin{eqnarray}\label{BH metric}
ds^2=-f(r)dt^{2}+\frac{dr^2}{f(r)}+r^{2}dx_{i}dx^{i},
\end{eqnarray}
with
\begin{eqnarray}
f(r)=\frac{r^2}{2\alpha}\left[1-\sqrt{1-\frac{4\alpha}{L^{2}}
\left(1-\frac{ML^{2}}{r^{d-1}}\right)}~\right],
\end{eqnarray}
where $M$ is a constant of integration relating to the black hole horizon by $r_{+}=(ML^{2})^{1/(d-1)}$, $\alpha$ is the Gauss-Bonnet coupling
constant and $L$ is the AdS radius. The Gauss-Bonnet correction  $\alpha$ has an upper bound called the Chern-Simons limit $\alpha=L^{2}/4$, and
a lower bound determined by the causality \cite{Buchel-Myers,Camanho-Edelstein}. In the limit $\alpha\rightarrow0$, (1) goes back to the
Schwarzschild AdS black hole.

Consider a $U(1)$ gauge field and the scalar field coupled via a generalized St\"{u}ckelberg Lagrangian \cite{Franco}
\begin{eqnarray}\label{System}
S=\int d^{d}x\sqrt{-g}\left[
-\frac{1}{4}F_{\mu\nu}F^{\mu\nu}-\frac{1}{2}\partial_{\mu}\tilde{\psi}\partial^{\mu}\tilde{\psi}
-\frac{1}{2}m^2\tilde{\psi}^2-\frac{1}{2}|\mathfrak{F}(\tilde{\psi})|(\partial_{\mu}p-A_{\mu})
(\partial^{\mu}p-A^{\mu}) \right] \ ,
\end{eqnarray}
with gauge symmetry $A_{\mu}\rightarrow
A_{\mu}+\partial_{\mu}\Lambda$ and $p\rightarrow p+\Lambda$.
$\mathfrak{F}$ is a general function of $\tilde{\psi}$ which has the
following form
\begin{eqnarray}\label{model}
\mathfrak{F}(\tilde{\psi})=\tilde{\psi}^{2}+c_{\gamma}\tilde{\psi}^{\gamma}+c_{4}\tilde{\psi}^{4},
\end{eqnarray}
with the model parameters $c_{\gamma}$, $\gamma$ and $c_{4}$. When $c_{\gamma}$ and $c_{4}$ are zero, it reduces to the model considered in
\cite{Gregory,Pan-Wang}.

Using the gauge freedom to fix $p=0$ and taking the ansatz $\psi\equiv\tilde{\psi}$, $A_{t}=\phi$ where $\psi$, $\phi$ are both real functions of
$r$ only, we can obtain the equations of motion
\begin{eqnarray}\label{Psi-Phi}
&&\psi^{\prime\prime}+\left(
\frac{f^\prime}{f}+\frac{d-2}{r}\right)\psi^\prime
+\frac{\phi^2}{2f^2}\mathfrak{F}^\prime(\psi)-\frac{m^2}{f}\psi=0\,,
\nonumber\\ &&
\phi^{\prime\prime}+\frac{d-2}{r}\phi^\prime-\frac{\mathfrak{F}(\psi)}{f}\phi=0~.
\end{eqnarray}
These two equations can be solved numerically by doing integration
from the horizon out to the infinity. At the asymptotic AdS boundary
($r\rightarrow\infty$), the solutions behave like
\begin{eqnarray}
\psi=\frac{\psi_{-}}{r^{\lambda_{-}}}+\frac{\psi_{+}}{r^{\lambda_{+}}}\,,\hspace{0.5cm}
\phi=\mu-\frac{\rho}{r^{d-3}}\,, \label{infinity}
\end{eqnarray}
with
\begin{eqnarray}
\lambda_\pm=\frac{1}{2}[(d-1)\pm\sqrt{(d-1)^{2}+4m^{2}L_{\rm
eff}^2}~]\,, \label{LambdaZF}
\end{eqnarray}
where $L_{\rm eff}^{2}=2\alpha/(1-\sqrt{1-4\alpha/L^2}~)$ is the effective asymptotic AdS scale \cite{Gregory,Pan-Wang}, $\mu$ and $\rho$ are
interpreted as the chemical potential and charge density in the dual field theory respectively. Notice that both of the falloffs are normalizable
for $\psi$, so one can impose boundary condition that either $\psi_{+}$ or $\psi_{-}$ vanishes \cite{HartnollPRL101,HartnollJHEP12}. For
simplicity, we will take $\psi_{-}=0$. Moreover, we will set $d=5$ and $m^2L^2=-3$ for concreteness. As a matter of fact, the other choices will
not qualitatively modify our results. Thus, the scalar condensate is now described by the operator $\langle{\cal O}_{+}\rangle=\psi_{+}$ and we
will discuss the condensate $\langle{\cal O}_{+}\rangle$ for fixed charge density.

\begin{figure}[H]
\includegraphics[scale=0.75]{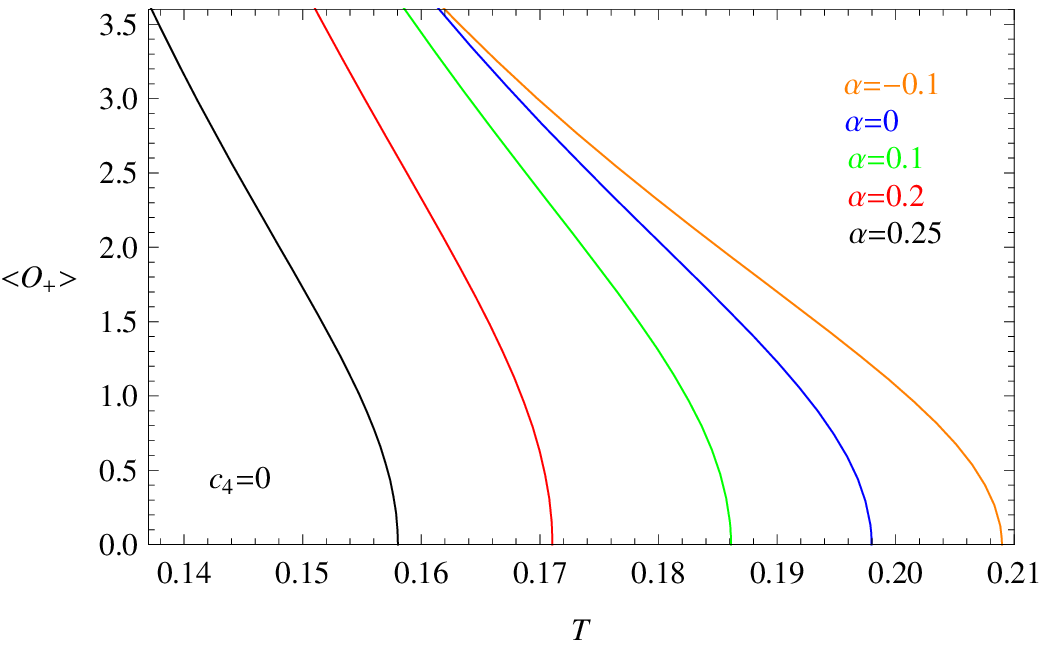}\hspace{0.2cm}%
\includegraphics[scale=0.75]{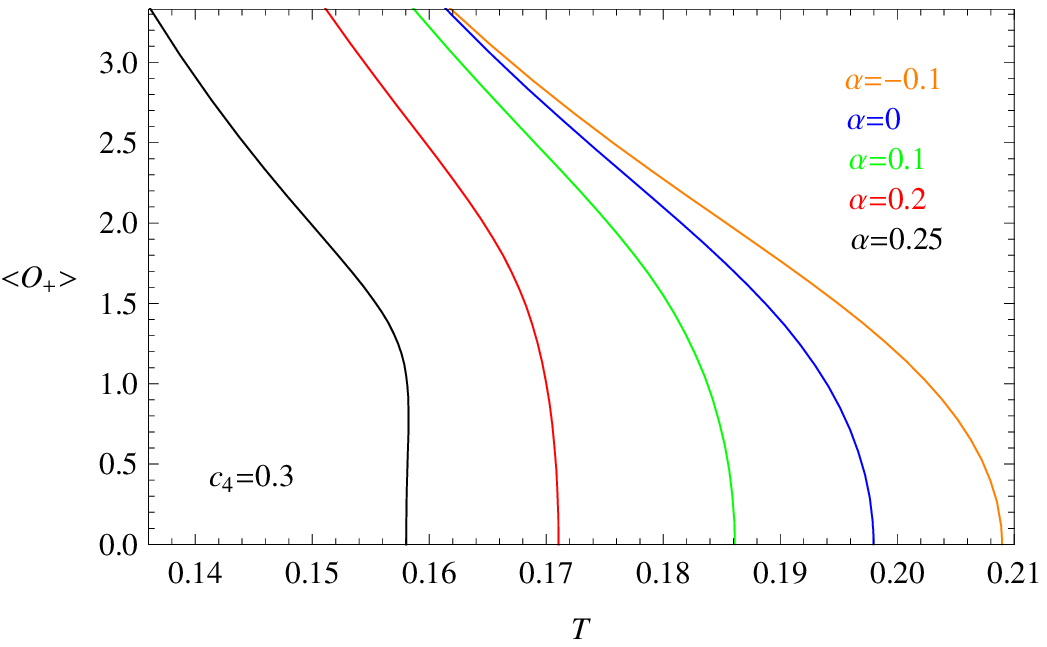}\\ \hspace{0.2cm}%
\includegraphics[scale=0.75]{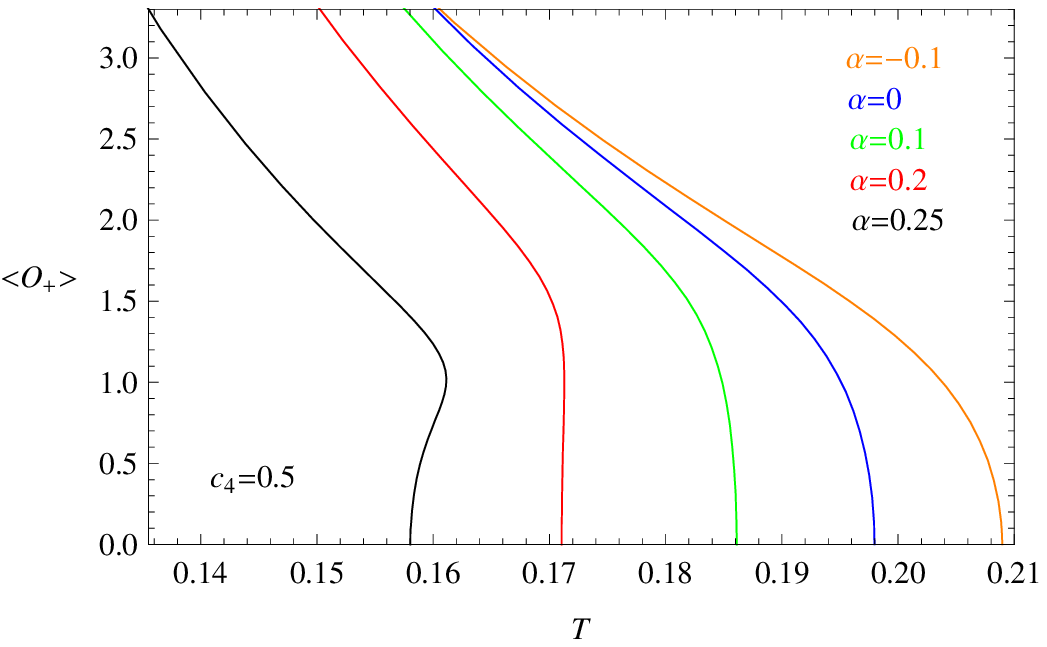}\vspace{0.0cm}
\includegraphics[scale=0.75]{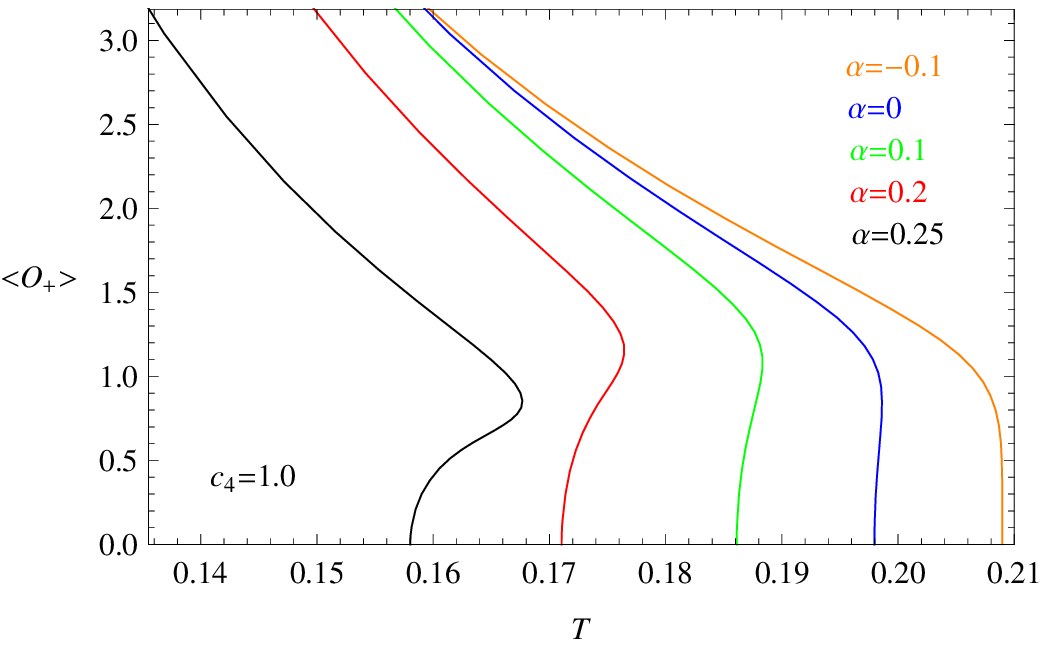}\\ \vspace{0.0cm}
\caption{\label{CondGBc4} (Color online) The condensate
$<\mathcal{O}_{+}>$ as a function of temperature with fixed values
$c_{4}$ ($c_{\gamma}=0$) for different values of $\alpha$, which
shows that a different value of $\alpha$ can separate the first- and
second-order behavior. The five lines in each panel from left to
right correspond to decreasing $\alpha$, i.e., $0.25$ (black), $0.2$
(red), $0.1$ (green), $0$ (blue) and $-0.1$ (orange).}
\end{figure}

We will investigate how the phase transition depends on the coefficients $c_{\gamma}$ and $c_{4}$. Due to the special interest in the case
$\mathfrak{F}(\psi)=\psi^{2}+c_{4}\psi^{4}$ \cite{Gubser-C-S-T,Gauntlett-Sonner}, in (\ref{model}) we are going to set $c_{\gamma}=0$ for the
moment and pay more attention on the influence of $c_4$ on the phase transition. Solving the equations of motion numerically, in
Fig.\ref{CondGBc4} we plot the condensate around the critical region for chosen values of $c_{4}$ and different Gauss-Bonnet constants. For
$0\leq c_{4}<0.3$, the transition is second order and the condensate approaches zero as $\langle{\cal O}_{+}\rangle\sim (T_{c}-T)^{\beta}$, with
mean field critical exponent $\beta=1/2$ for all values of $\alpha$. For $c_{4}\geq 1$, we observe that $\langle{\cal O}_{+}\rangle$ becomes
multivalued near the critical temperature and the condensate does not drop to zero continuously at the critical temperature. This behavior keeps
for all values of $\alpha$. The analogous phenomenon holds as well when we consider the $\langle{\cal O}_{-}\rangle$ condensate but with
different $c_4$ range. In \cite{Franco}, for the case without Gauss-Bonnet constant, it was argued that the behavior for $c_4\geq 1$ indicates
that the phase transition changes from the second order to the first order at $c_4=1$. Here we find that the Gauss-Bonnet constant does not alter
the result when $c_4\geq 1$. Choosing $c_{4}\in[0.3,1.0]$, we observe in Fig. \ref{CondGBc4} that for fixed $c_4$, the transition point of the
phase transition from the second order to the first order appears easily for the bigger value of $\alpha$.  In table \ref{critical value}, we
list the critical value of $\alpha_{c}$  separating the second order and the first order phase transitions for selected $c_4$ within the range
$[0.3,1.0]$. Since we concentrate on the five-dimensional spacetime, the separation point between the second and the first order phase
transitions does not coincide with that in four dimensions, for example $(c_4=0.9,\alpha_c=0)$ here and $(c_4=1, \alpha=0)$ in \cite{Franco}.
With the increase of $c_4$, we see that $\alpha_c$ becomes smaller. Thus we find that when $c_{4}\in[0.3,1.0]$, not only $c_4$ but also the
Gauss-Bonnet constant can tune the order of the phase transition. The Gauss-Bonnet constant provides richer physics in the phase transition.

\begin{table}[ht]
\caption{\label{critical value} The critical value of $\alpha_{c}$
which can separate the first- and second-order behavior for
different $\mathfrak{F}(\psi)=\psi^{2}+c_{4}\psi^{4}$. }
\begin{tabular}{|c|c|c|c|c|c|}
         \hline
$~~c_{4}~~$ &~~0.3~~&~~0.5~~&~~0.7~~&~~0.9~~&~~1.0~~
          \\
        \hline
~~$\alpha_{c}$~~ &~~0.25~~&~~0.2~~&~~0.1~~&~~0.0~~&~~-0.1~~
          \\
        \hline
\end{tabular}
\end{table}

\begin{figure}[H]
\includegraphics[scale=0.75]{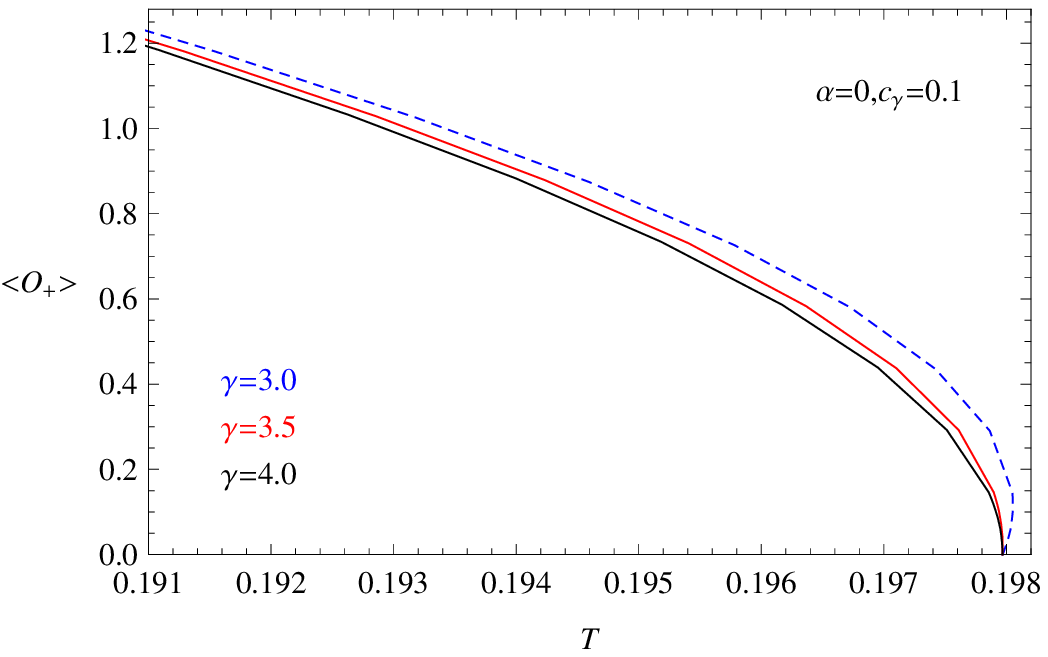}\hspace{0.2cm}%
\includegraphics[scale=0.75]{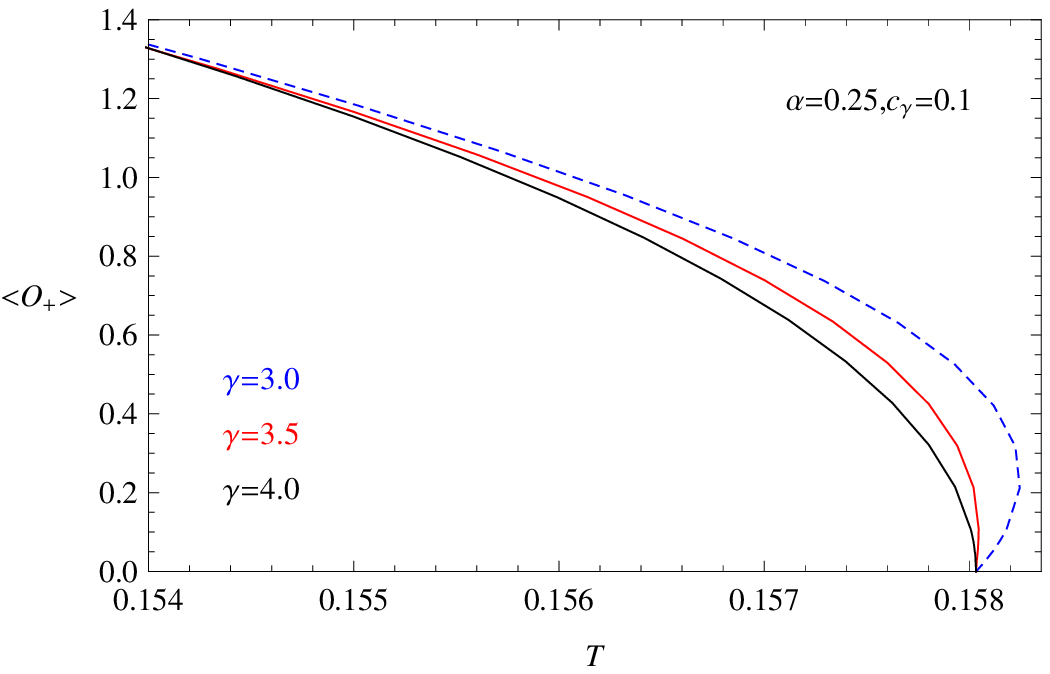}\\ \vspace{0.0cm}
\includegraphics[scale=0.75]{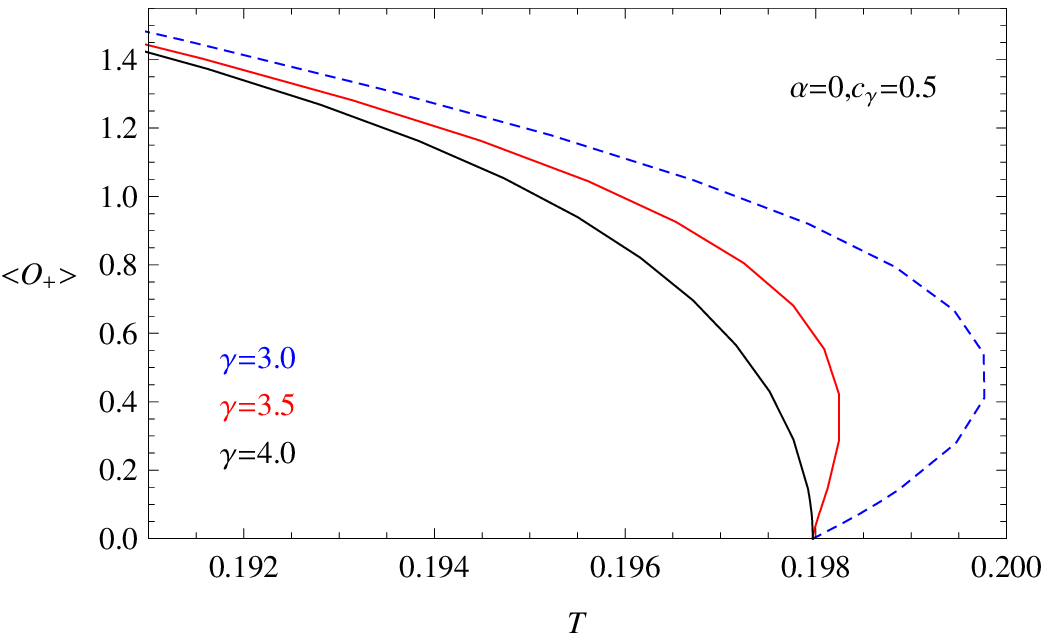}\hspace{0.2cm}%
\includegraphics[scale=0.75]{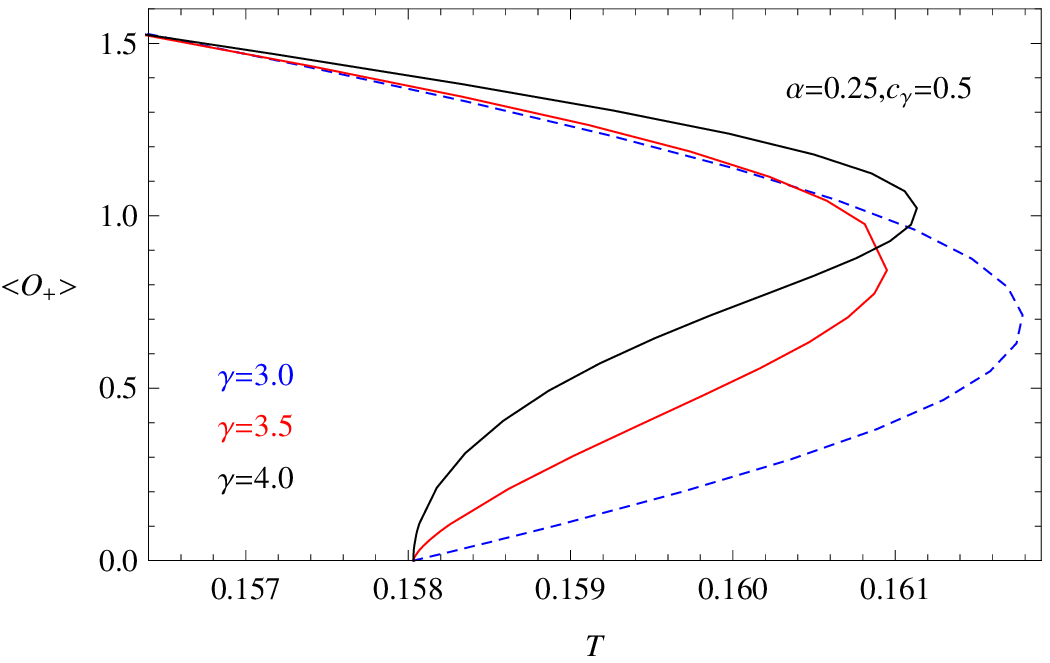}\\ \vspace{0.0cm}
\caption{\label{CondGBcLambda} (Color online) The condensate
$<\mathcal{O}_{+}>$ as a function of temperature with fixed values
$c_{\gamma}$ and $\alpha$ for different values of $\gamma$, which
shows that the type of phase transition depends on the Gauss-Bonnet
correction term $\alpha$, model parameters $\gamma$ and
$c_{\gamma}$. The three lines in each panel from left to right
correspond to decreasing $\gamma$, i.e., $4.0$ (black), $3.5$ (red)
and $3.0$ (blue and dashed).}
\end{figure}

We also have the interest to see the influence of $c_{\gamma}$ on the phase transition. We will concentrate on $\gamma$ in the range
$3\leq\gamma\leq4$. At this moment we set $c_4=0$. In Fig. \ref{CondGBcLambda} we exhibit the condensate of $<\mathcal{O}_{+}>$ for selected
values of $c_{\gamma}$, $\alpha$ and changing $\gamma$. We see that for fixed small values of $\alpha$ and $c_{\gamma}$, it is more possible to
observe the appearance of the first order phase transition when $\gamma$ is smaller. The $\alpha$ influence on the order of the phase transition
is also exhibited. Bigger $\alpha$ will cause the first order phase transition to appear easier for fixed $c_{\gamma}$ and $\gamma$. This is
consistent with the case for nonzero $c_4$ but zero $c_{\gamma}$. The effect of $c_{\gamma}$ is similar to the Gauss-Bonnet constant, bigger
$c_{\gamma}$ brings the first order phase transition easier for fixed $\gamma$ and $\alpha$.

\begin{figure}[H]
\includegraphics[scale=0.75]{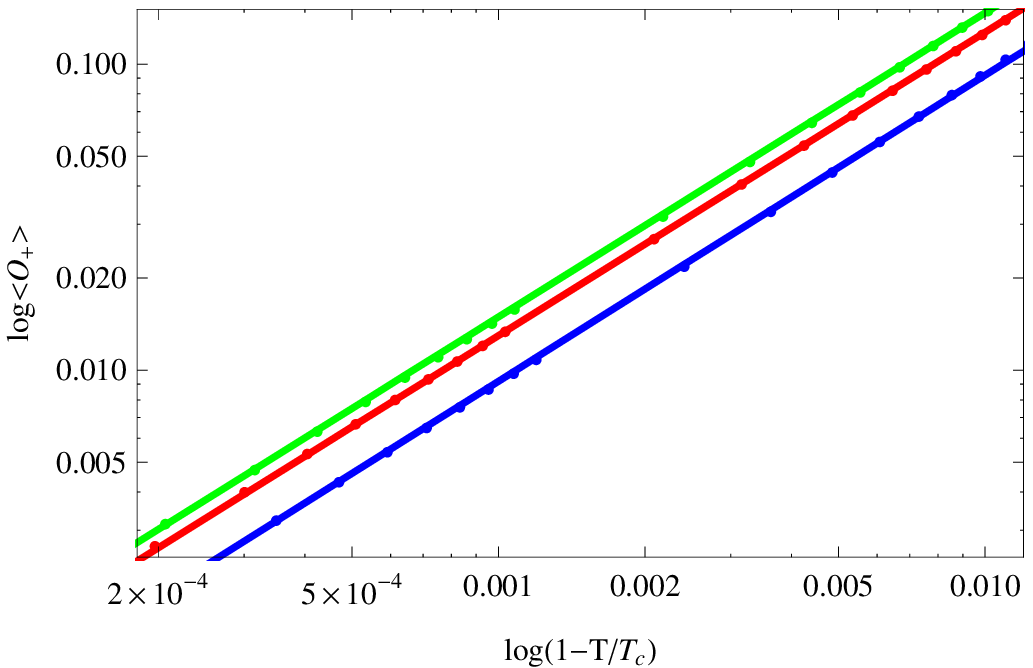}\hspace{0.2cm}%
\includegraphics[scale=0.75]{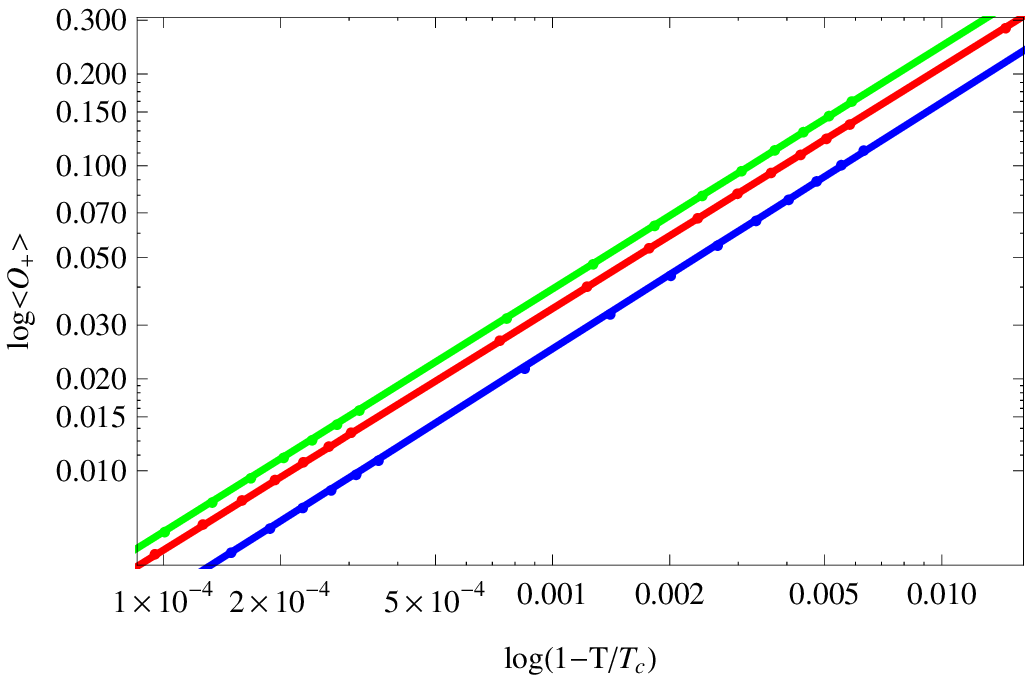}\\ \vspace{0.0cm}
\includegraphics[scale=0.75]{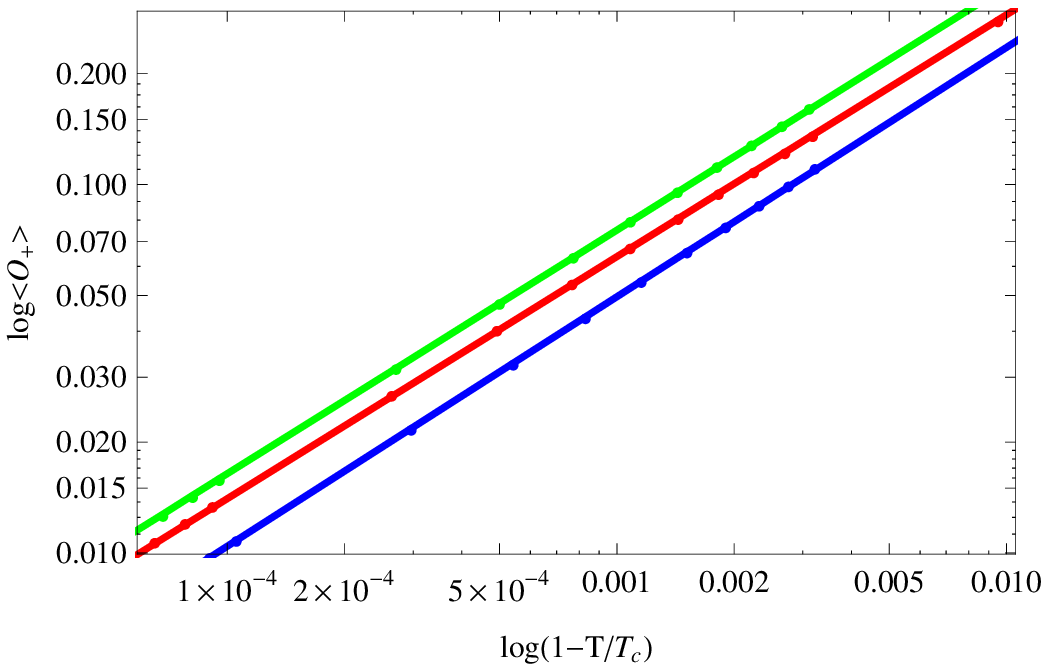}\hspace{0.2cm}%
\includegraphics[scale=0.75]{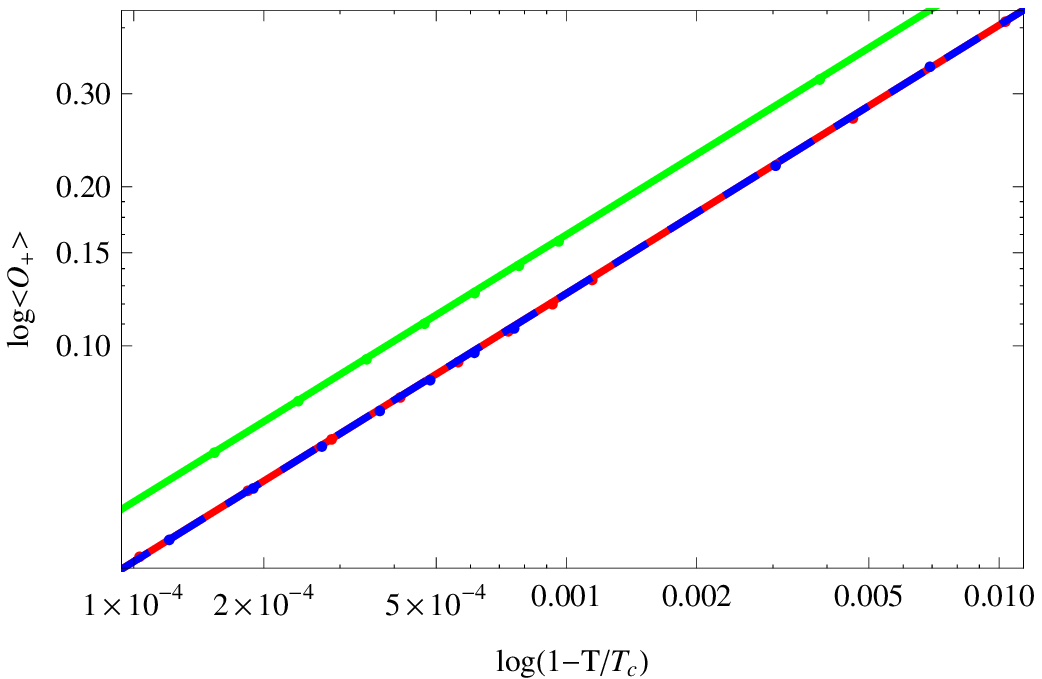}\\ \vspace{0.0cm}
\caption{\label{CondGBNH} (color online) The condensate
$<\mathcal{O}_{+}>$ vs $1-T/T_{c}$ in logarithmic scale with
different values of $\alpha$ for $\gamma=3.0$ (top and left), $3.25$
(top and right), $3.5$ (bottom and left) and $4.0$ (bottom and
right). The three lines in each panel from bottom to top correspond
to $\alpha=0.25$ (blue), $-0.1$ (red) and $0.1$ (green). These
panels show that the slope is independent of $\alpha$ but sensitive
to $\gamma$.}
\end{figure}

In addition to showing that the gravity duals can lead to both first
and second order phase transitions, it is also of great interest to
examine in the second order phase transition  whether different
choices of $c_{\gamma}, c_4, \gamma, \alpha$ result in different
critical exponents from the prediction of the mean field. When
$c_{\gamma}=0$ and $c_4=0$, the critical exponent was shown in
agreement with the mean field value $\beta=1/2$ in the
Einstein-Gauss-Bonnet gravity \cite{Gregory,Pan-Wang}. An
interesting behavior arises for $c_{\gamma}<0$. In Fig.
\ref{CondGBNH}, for $c_{\gamma}=-1$ and $c_{4}=1/2$, we present the
condensate $<\mathcal{O}_{+}>$ as a function of $1-T/T_{c}$ in
logarithmic scale with different values of $\alpha$ for choosing
$\gamma=3.0$,  $3.25$, $3.5$ and $4.0$ respectively.  Three lines in
each panel from the bottom to the top correspond to $\alpha=0.25$,
$-0.1$ and $0.1$. We see that the slope is independent of the
Gauss-Bonnet correction term $\alpha$ but sensitive to the model
parameter $\gamma$. Further analysis shows that, near the critical
temperature $T_{c}$, the critical exponent $\beta\simeq1.00$ for
$\gamma=3.0$, $\beta\simeq0.79$ for $\gamma=3.25$, $\beta\simeq0.67$
for $\gamma=3.5$ and $\beta\simeq0.50$ for $\gamma=4.0$, which is
independent of $\alpha$! The relation between the critical exponent
$\beta$ and the parameter $\gamma$ can be expressed as
\begin{eqnarray}\label{critial-exponent}
\beta\cong\frac{1}{\gamma-2}.
\end{eqnarray}
This behavior is consistent with that seen for the AdS$_{4}$ black hole with the scalar mass $m^{2}L^{2}=-2$ \cite{Franco,Aprile-Russo}, which
tells us that the critical exponent $\beta$  depends only on the model parameter $\gamma$ but is independent of the scalar mass and the
background spacetimes.

\begin{figure}[H]
\includegraphics[scale=0.75]{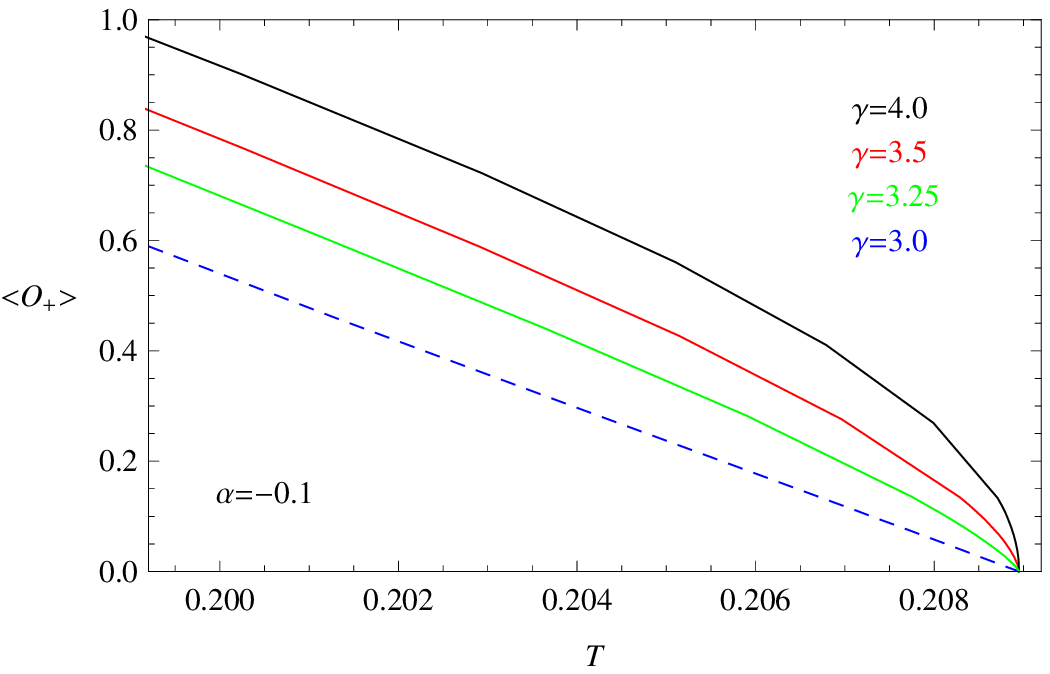}\hspace{0.2cm}%
\includegraphics[scale=0.75]{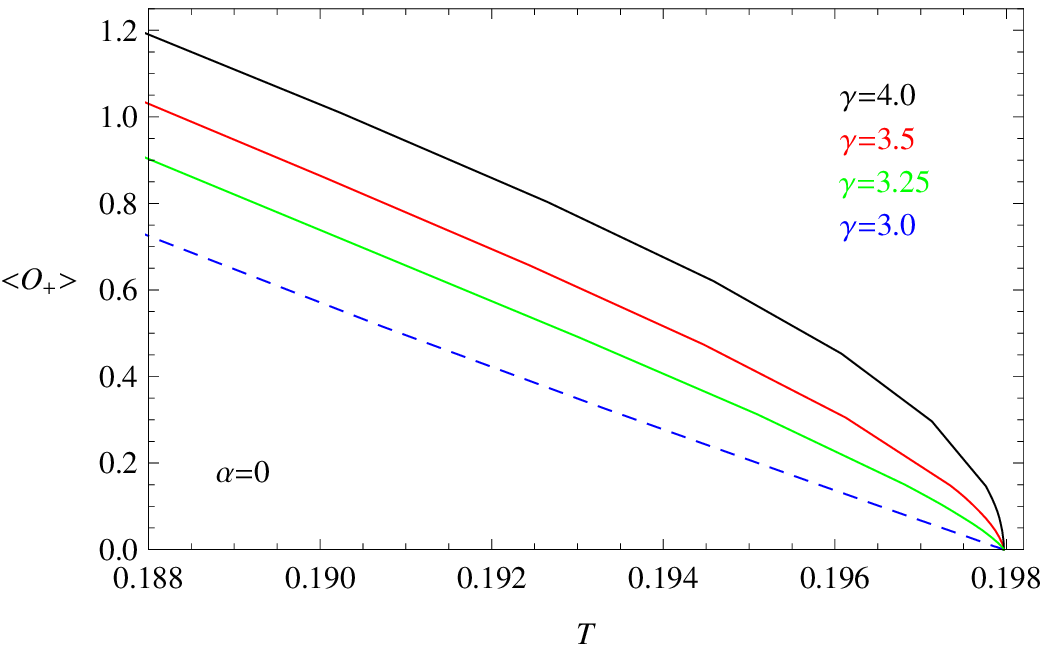}\\ \vspace{0.0cm}
\includegraphics[scale=0.75]{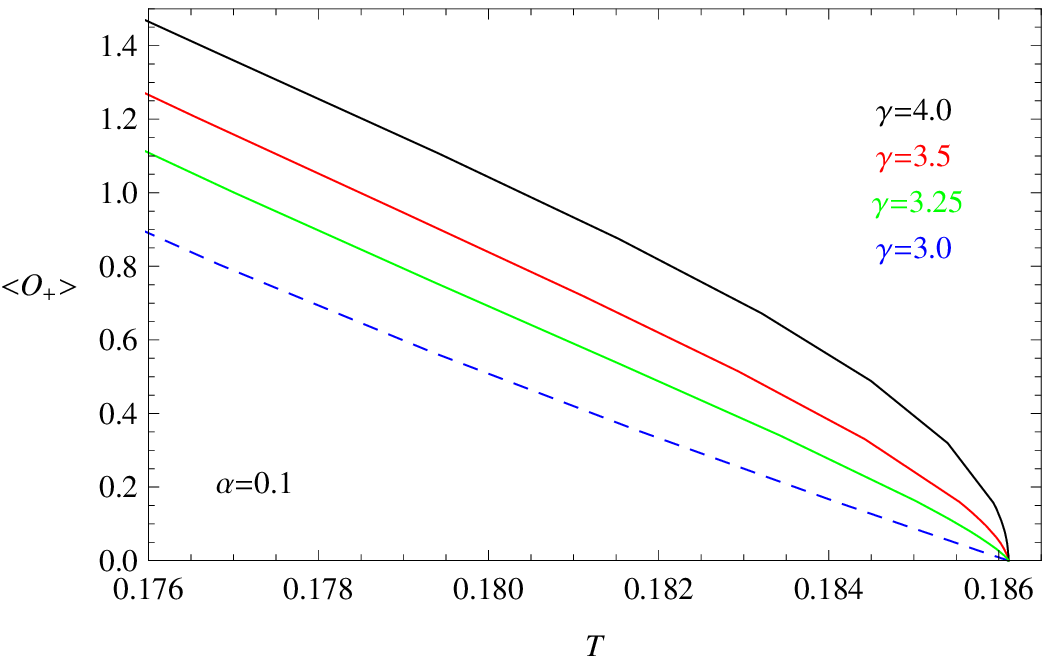}\hspace{0.2cm}%
\includegraphics[scale=0.75]{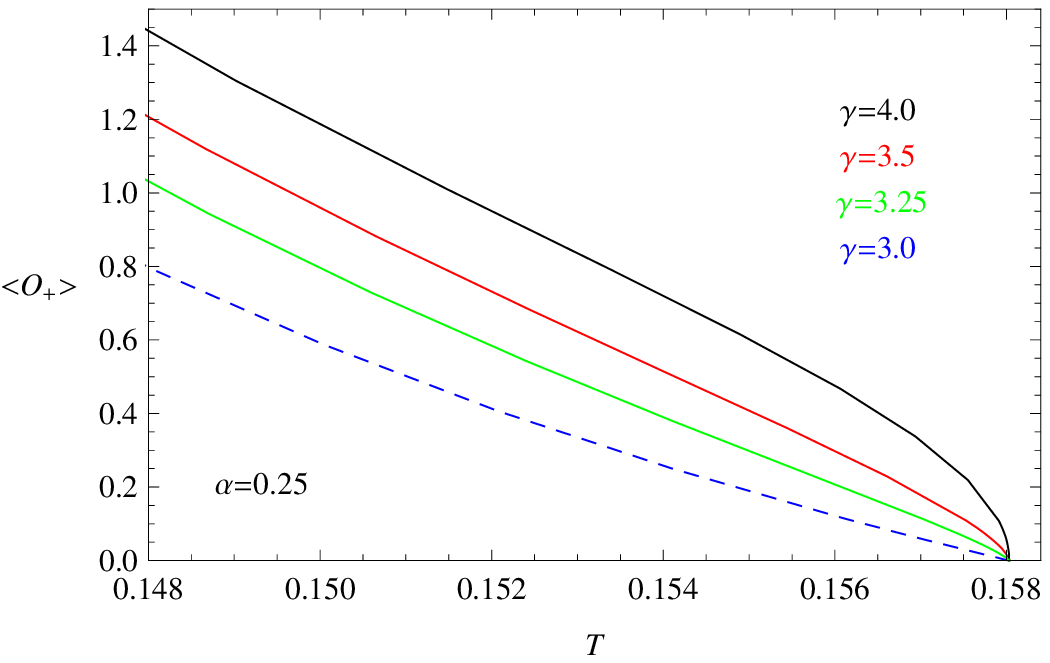}\\ \vspace{0.0cm}
\caption{\label{CondGBTc} (color online) The condensate
$<\mathcal{O}_{+}>$ as a function of temperature with fixed values
$\alpha$ for different model parameters $\gamma$. The four lines in
each panel from bottom to top correspond to increasing $\gamma$,
i.e., $3.0$ (blue and dashed), $3.25$ (green), $3.5$ (red) and $4.0$
(black). These panels show that the critical temperature $T_{c}$ is
independent of $\gamma$ but depends on $\alpha$.}
\end{figure}

In Fig. \ref{CondGBTc} we present the condensate as a function of temperature for $c_{\gamma}=-1$ and $c_{4}=1/2$. It exhibits that the critical
temperature $T_{c}$ is independent of the model parameter $\gamma$ but depends on the Gauss-Bonnet correction term $\alpha$, i.e., $T_{c}=0.209$
for $\alpha=-0.1$, $T_{c}=0.198$ for $\alpha=0$, $T_{c}=0.186$ for $\alpha=0.1$ and $T_{c}=0.158$ for $\alpha=0.25$, which shows that the
positive Gauss-Bonnet correction will suppress the condensation but the negative one will enhance it \cite{Ge-Wang}. In fact, we note that the
critical temperature $T_{c}$ is not sensitive to the coefficients $c_{\gamma}$ and $c_{4}$ in the general function $\mathfrak{F}(\psi)$ from
Figs. \ref{CondGBc4}, \ref{CondGBcLambda} and \ref{CondGBTc} for the phase transition of the second order. Thus, we point out that the critical
temperature $T_{c}$ depends on the background spacetimes but not on the model of $\mathfrak{F}$.

\section{Conductivity}

Now we jump to investigate the influence of the Gauss-Bonnet correction term $\alpha$, model parameters $\gamma$, $c_{\gamma}$ and $c_{4}$ in
$\mathfrak{F}(\psi)$ on the conductivity.

In order to calculate the conductivity, we consider the perturbed Maxwell field $\delta A_{x}=A_{x}(r)e^{-i\omega t}dx$. The equation of motion
for $\delta A_{x}$ reads
\begin{eqnarray}
A_{x}^{\prime\prime}+\left(\frac{f^\prime}{f}+\frac{d-4}{r}\right)A_{x}^\prime
+\left[\frac{\omega^2}{f^2}-\frac{\mathfrak{F}(\psi)}{f}\right]A_{x}=0
\; . \label{Maxwell Equation}
\end{eqnarray}
We still restrict our study to $d=5$ in order to avoid the
complicated behavior in the gauge field falloff in dimensions higher
than five. We solve the above equation by imposing the ingoing
boundary condition $A_{x}(r)\sim f(r)^{-\frac{i\omega}{4r_+}}$ near
the horizon. In the asymptotic AdS region,
$A_{x}=A^{(0)}+\frac{A^{(2)}}{r^2} +\frac{A^{(0)}\omega^2 L_{\rm
eff}^4}{2} \frac{\log\Lambda r}{r^2}$. It should be noted
that the appearance of the arbitrary integration constant,
$\Lambda$, leads to a logarithmic divergence in the retarded 
Green's function $G^{R}$ which gives the conductivity $\sigma$. In
order to remove the divergent term $\log\Lambda r$, we add a
boundary counter term in the gravity action as suggested in \cite{Taylor}, which
can specify the renormalization scale when regulating
the action \cite{HorowitzPRD78}. With this appropriate boundary
counter term to cancel the logarithmic divergence, we can express
the conductivity as \cite{HorowitzPRD78,Gregory,Pan-Wang}
\begin{eqnarray}\label{GBConductivity}
\sigma=\frac{2A^{(2)}}{i\omega A^{(0)}}+\frac{i\omega}{2} \ .
\end{eqnarray}
For the general forms of function $\mathfrak{F}(\psi)=\psi^{2}+c_{\gamma}\psi^{\gamma}$ (setting $c_{4}=0$ for clarity), one can obtain the
conductivity by solving the Maxwell equation numerically. We will focus on the case for the fixed scalar mass $m^{2}L^{2}=-3$ in our discussion.

\begin{figure}[H]
\includegraphics[scale=0.5]{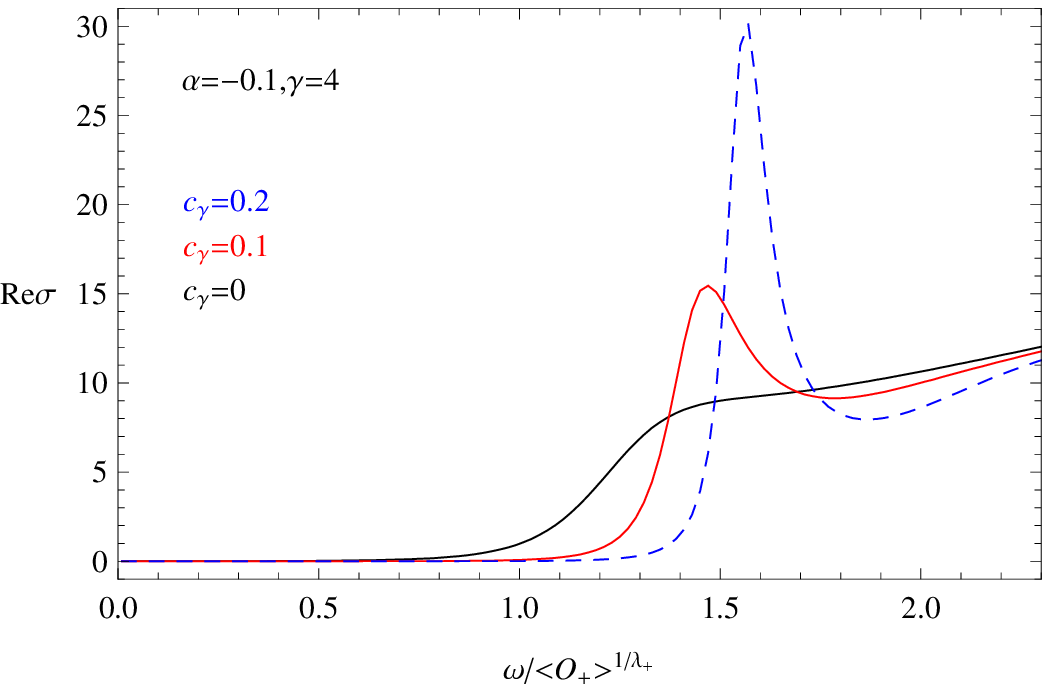}\hspace{0.2cm}%
\includegraphics[scale=0.5]{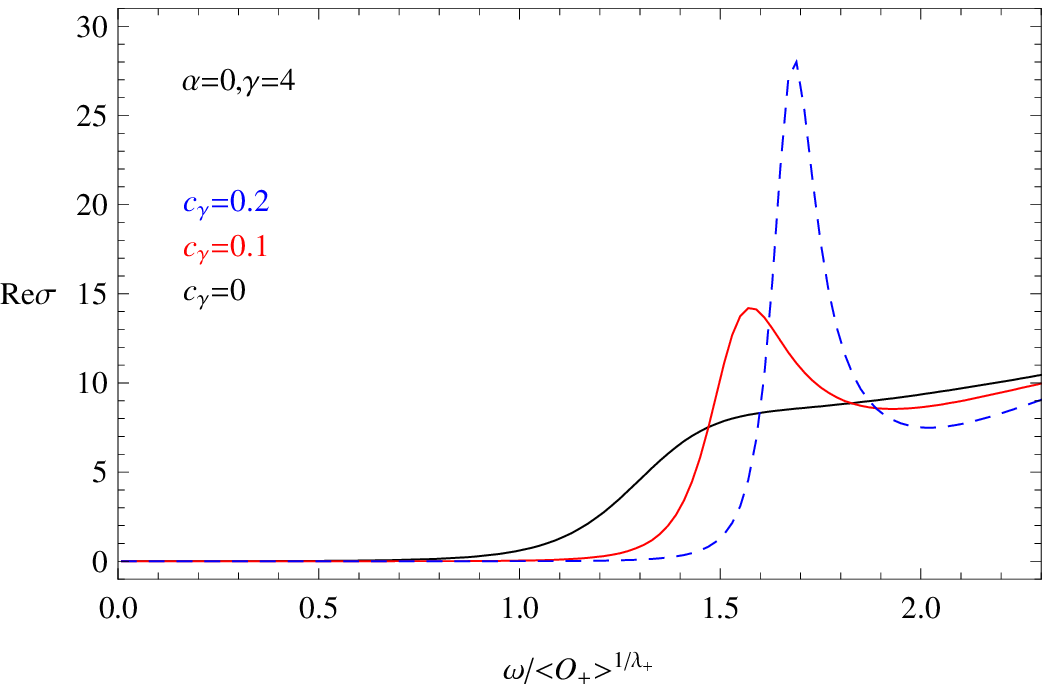}\hspace{0.2cm}%
\includegraphics[scale=0.5]{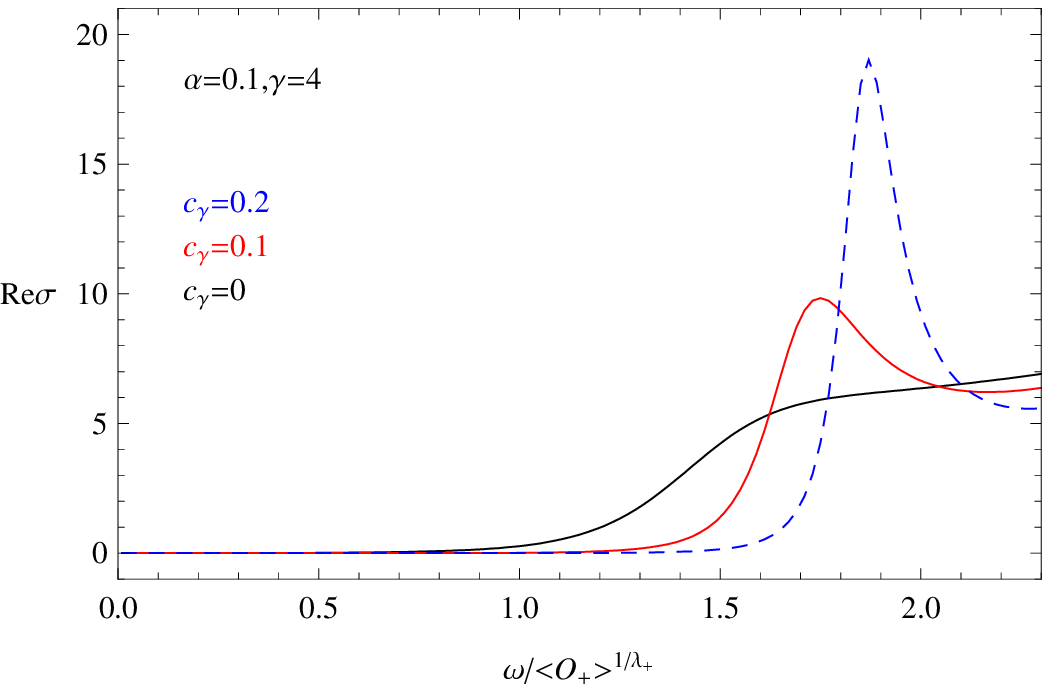}\\ \vspace{0.0cm}
\includegraphics[scale=0.5]{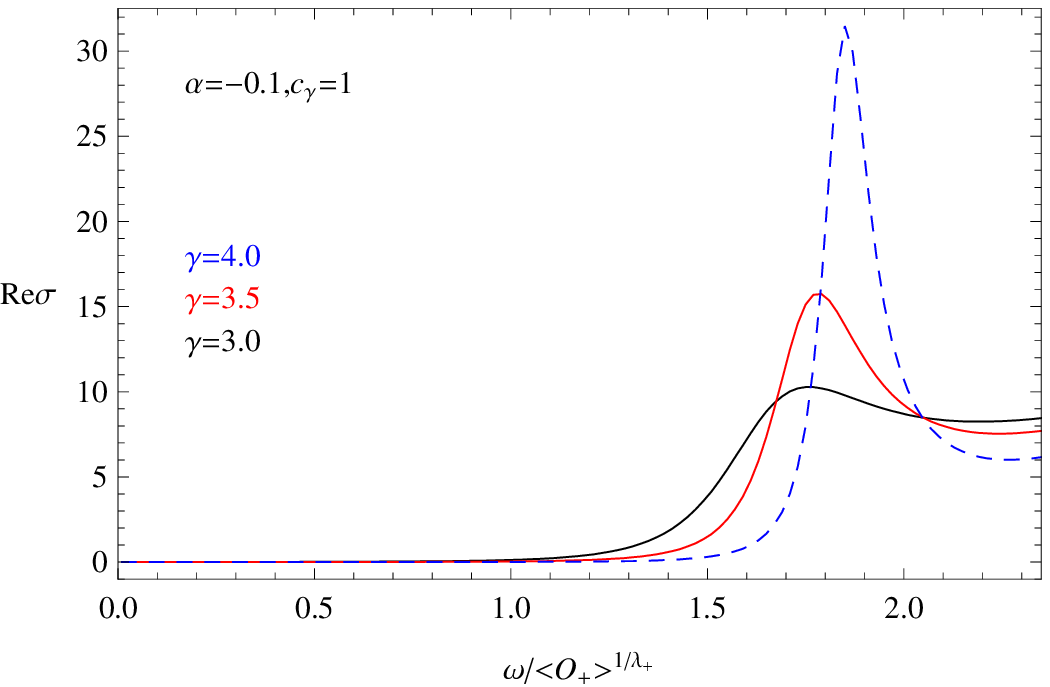}\hspace{0.2cm}%
\includegraphics[scale=0.5]{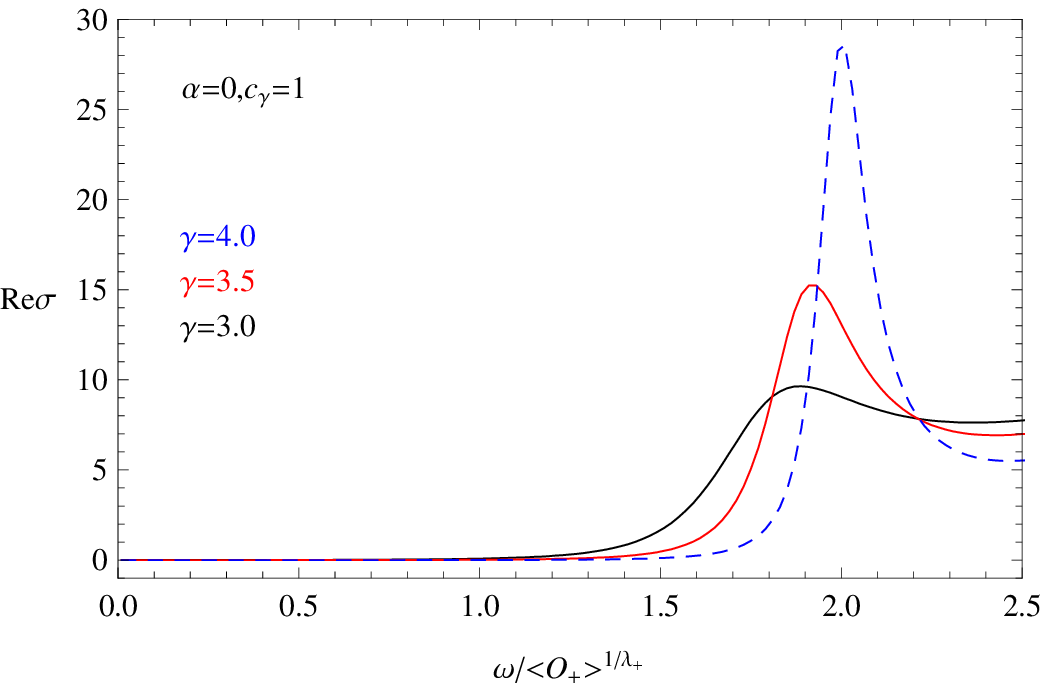}\hspace{0.2cm}%
\includegraphics[scale=0.5]{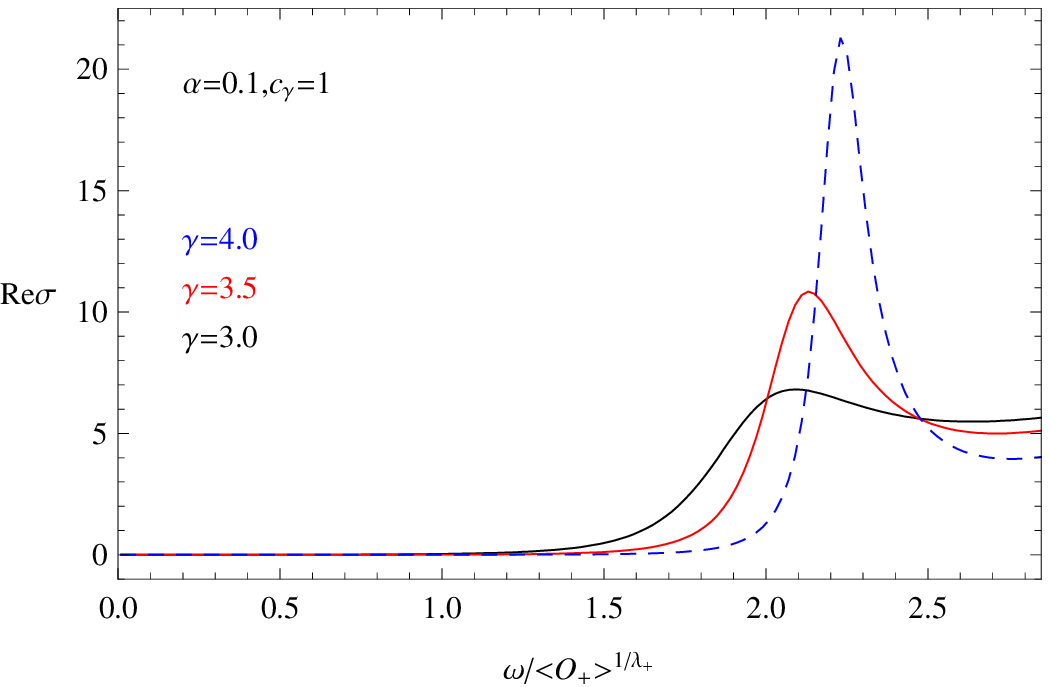}\\ \vspace{0.0cm}
\caption{\label{c4-Gamma-Conductivity} (color online) The real part
of the conductivity for ($3+1$)-dimensional Gauss-Bonnet
superconductors with different values of $\gamma$, $c_{\gamma}$ and
$\alpha$.}
\end{figure}

In the top-three panels in Fig. \ref{c4-Gamma-Conductivity} we
have plotted $Re\sigma(\omega)$ by fixing $\gamma=4$ for different values of
$c_{\gamma}$ and $\alpha$. It clearly shows that for fixed $\alpha$
the coherence peak gradually becomes stronger and narrower with the
increase of $c_{\gamma}$. This is consistent with the result of
$c_4$  observed in \cite{Franco}, which indicates that $c_{\gamma}$
controls the magnitude of the fluctuations of the condensate. For
bigger $c_{\gamma}$ we have already seen that the fluctuations are
strong enough to induce the first order phase transition.
In the bottom-three panels in Fig. \ref{c4-Gamma-Conductivity}
we show the result by fixing $c_{\gamma}=1$ and varying
$\gamma$ with selected $\alpha$ at $T/T_{c}\simeq0.7$. We observe
that with the increase of $\gamma$ the coherence peak increases for
the same $\alpha$, which shows that the model parameter $\gamma$
also controls the magnitude of the fluctuations. From Fig.
\ref{c4-Gamma-Conductivity}, we also learn that for the selected
$c_{\gamma}$ and $\gamma$, the coherence peak becomes lower with the
larger $\alpha$, which shows that the fluctuations in holographic
superconductors will be suppressed by the higher order curvature
corrections.

\begin{figure}[H]
\includegraphics[scale=0.5]{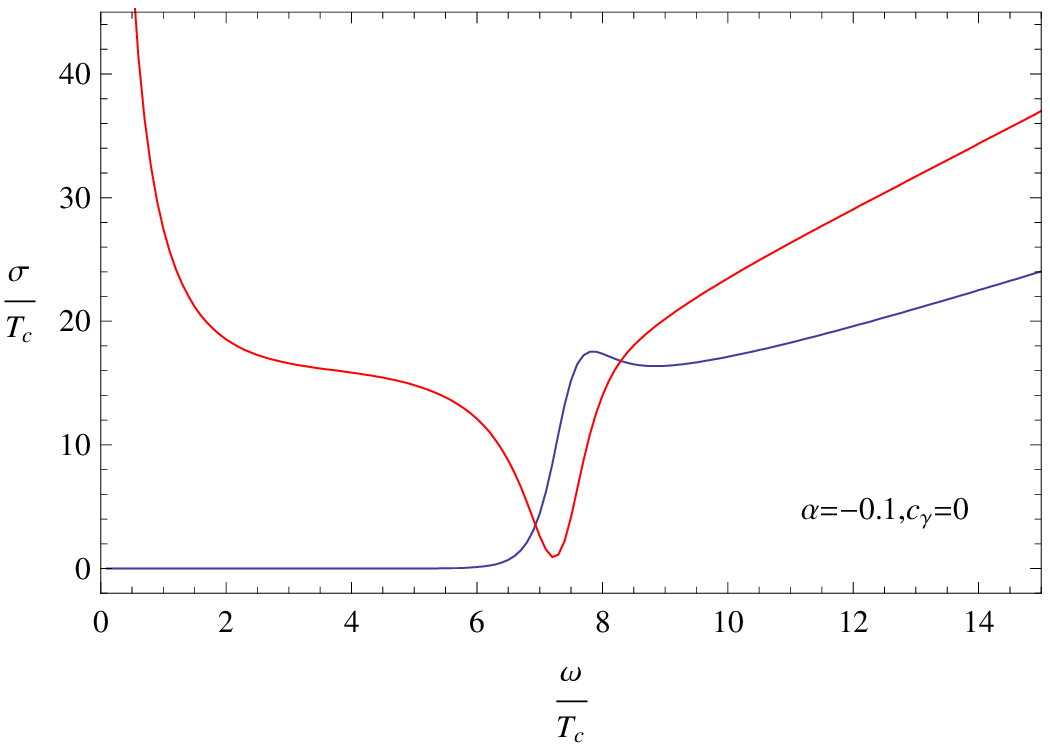}\hspace{0.2cm}%
\includegraphics[scale=0.5]{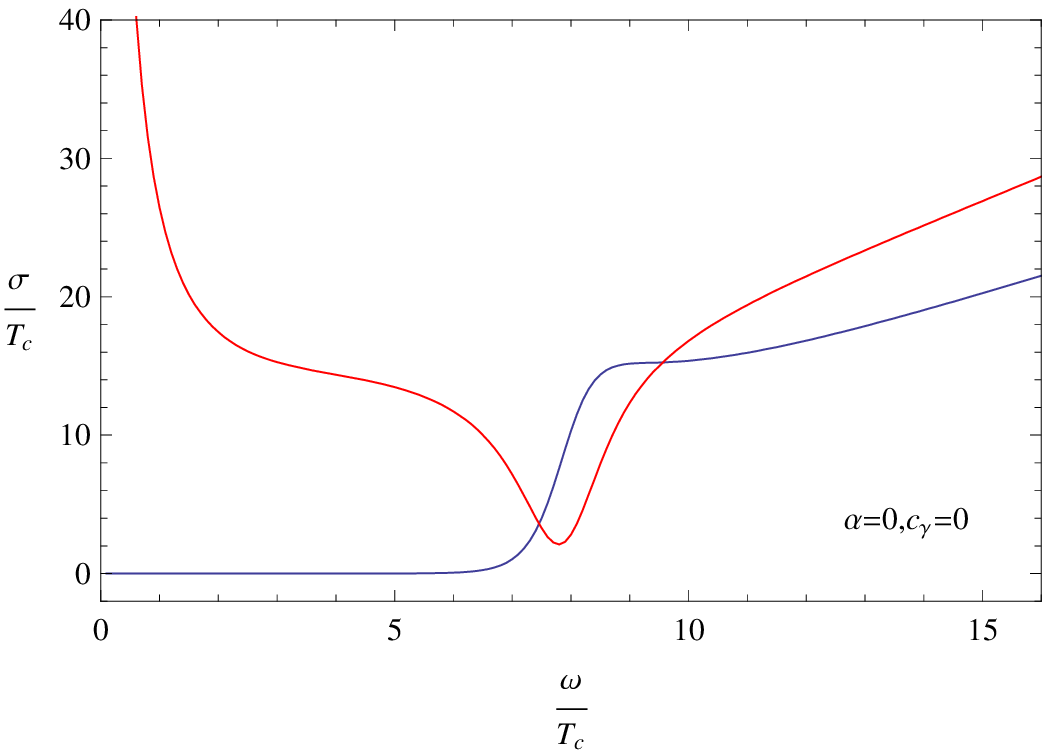}\hspace{0.2cm}%
\includegraphics[scale=0.5]{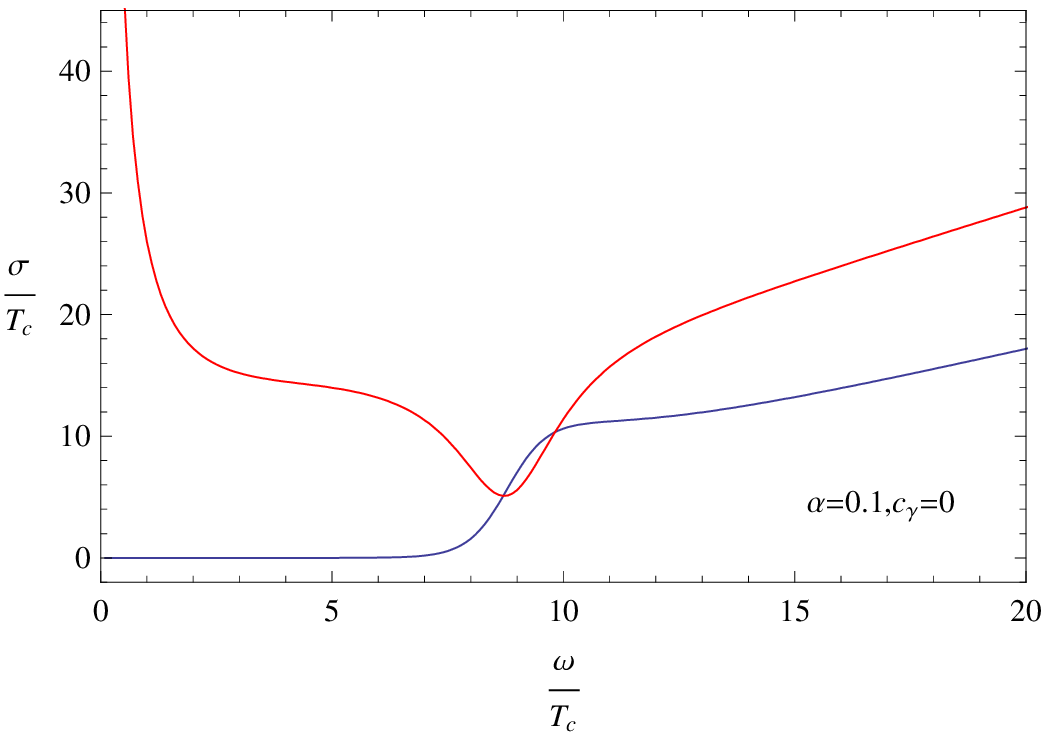}\\ \vspace{0.0cm}
\includegraphics[scale=0.5]{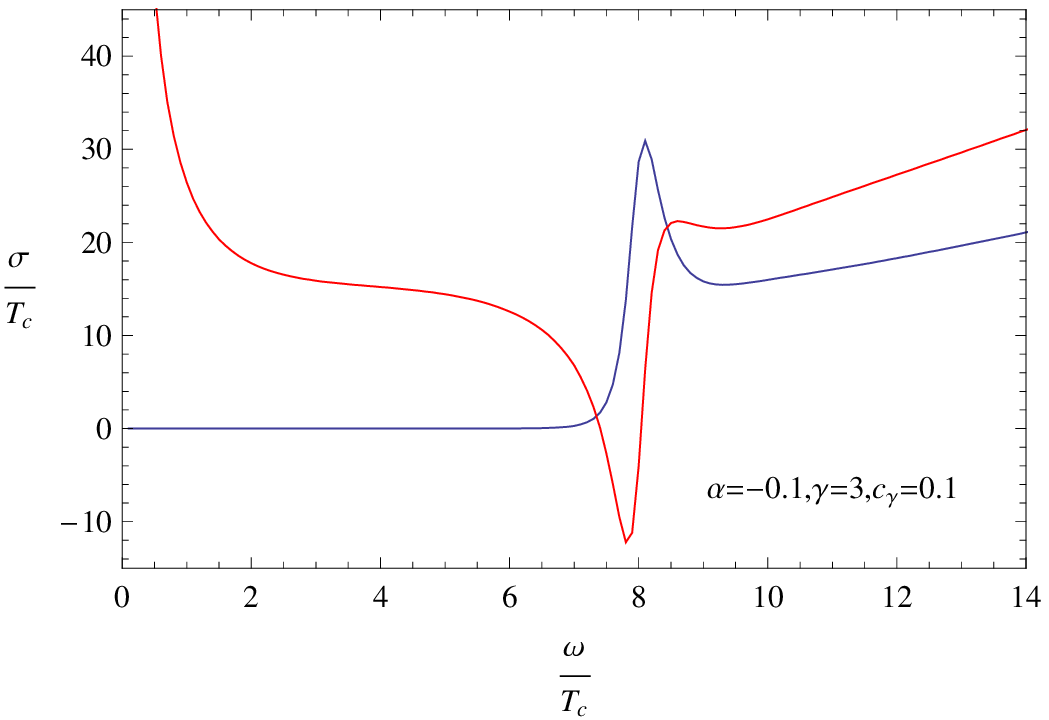}\hspace{0.2cm}%
\includegraphics[scale=0.5]{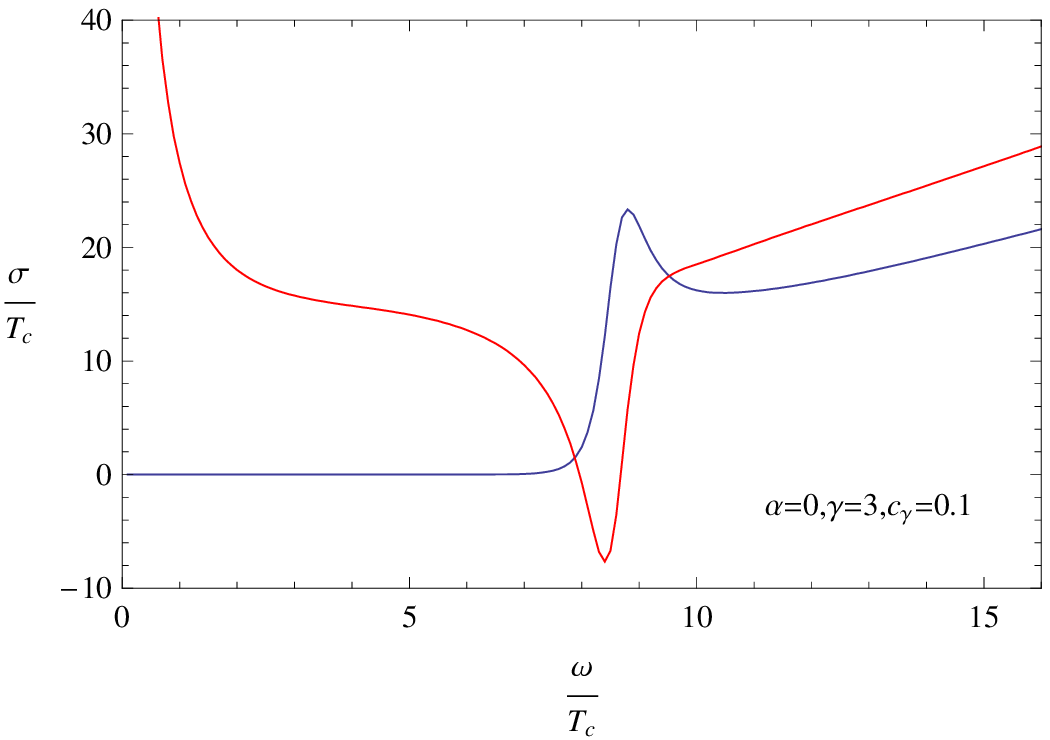}\hspace{0.2cm}%
\includegraphics[scale=0.5]{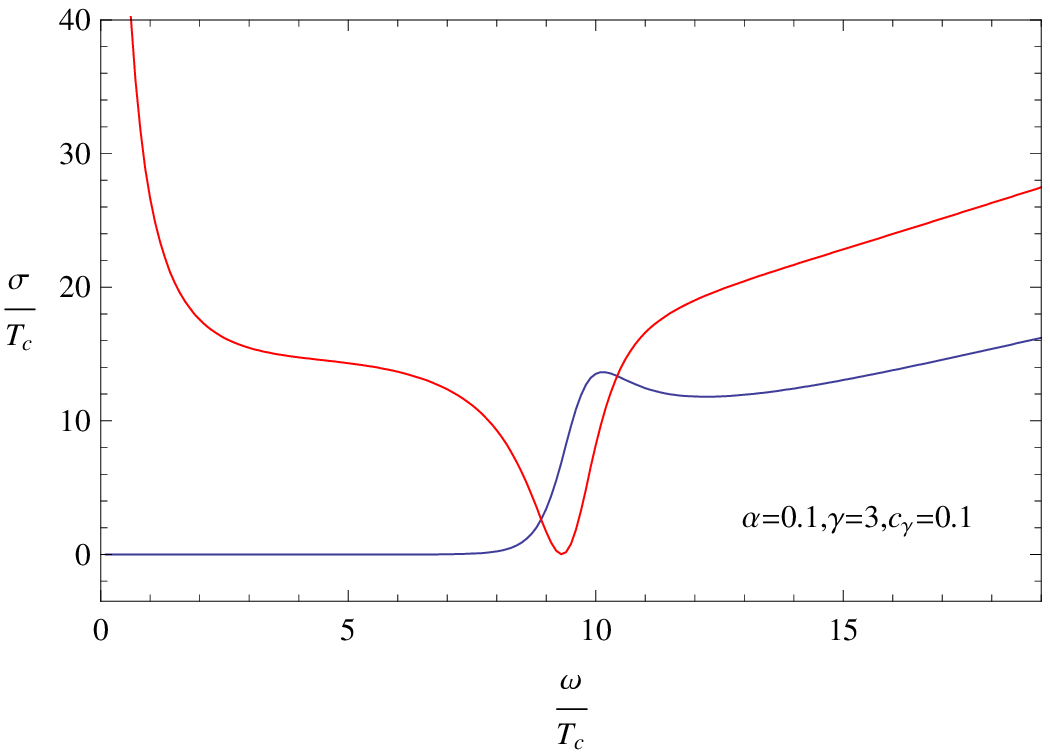}\\ \vspace{0.0cm}
\includegraphics[scale=0.5]{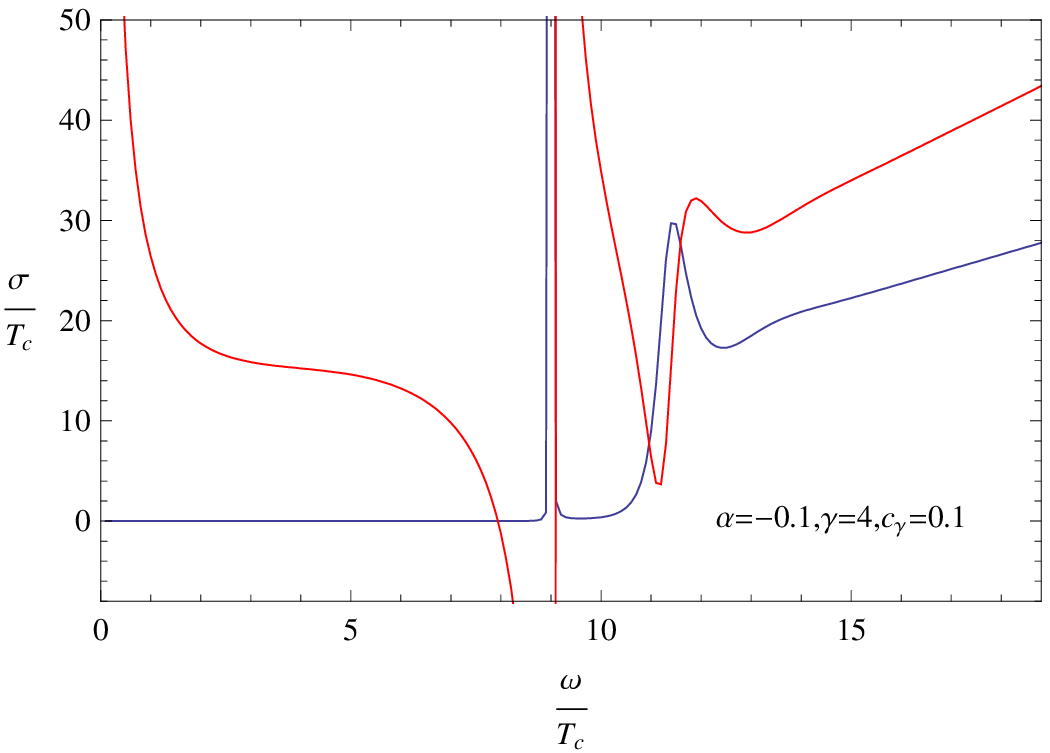}\hspace{0.2cm}%
\includegraphics[scale=0.5]{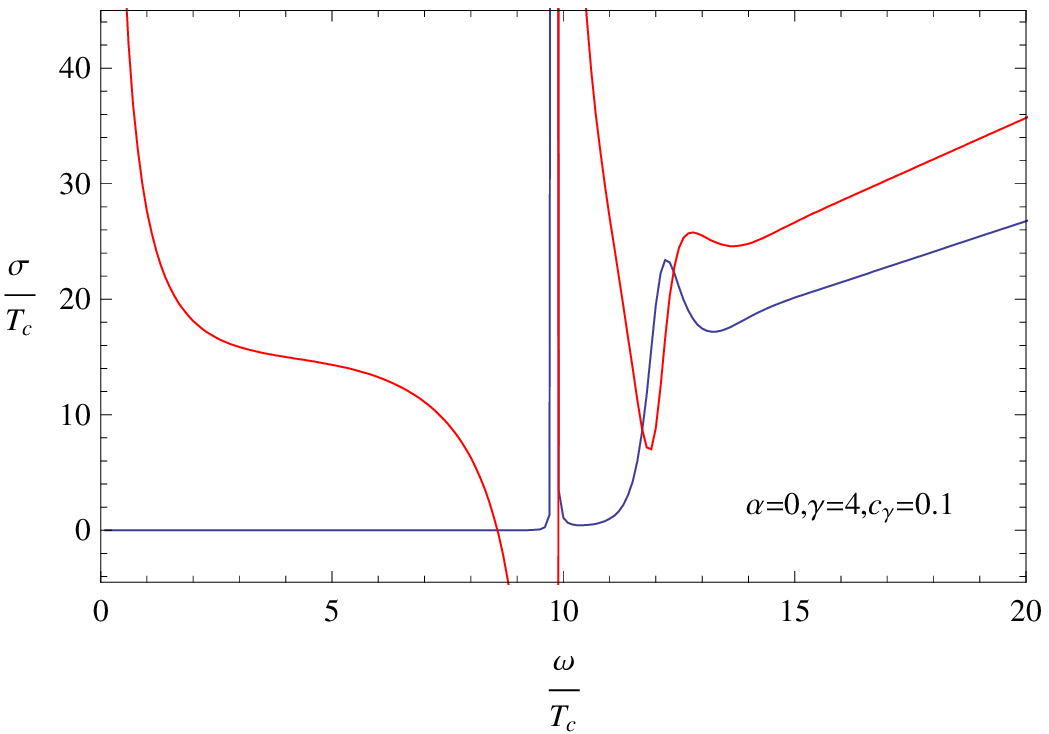}\hspace{0.2cm}%
\includegraphics[scale=0.5]{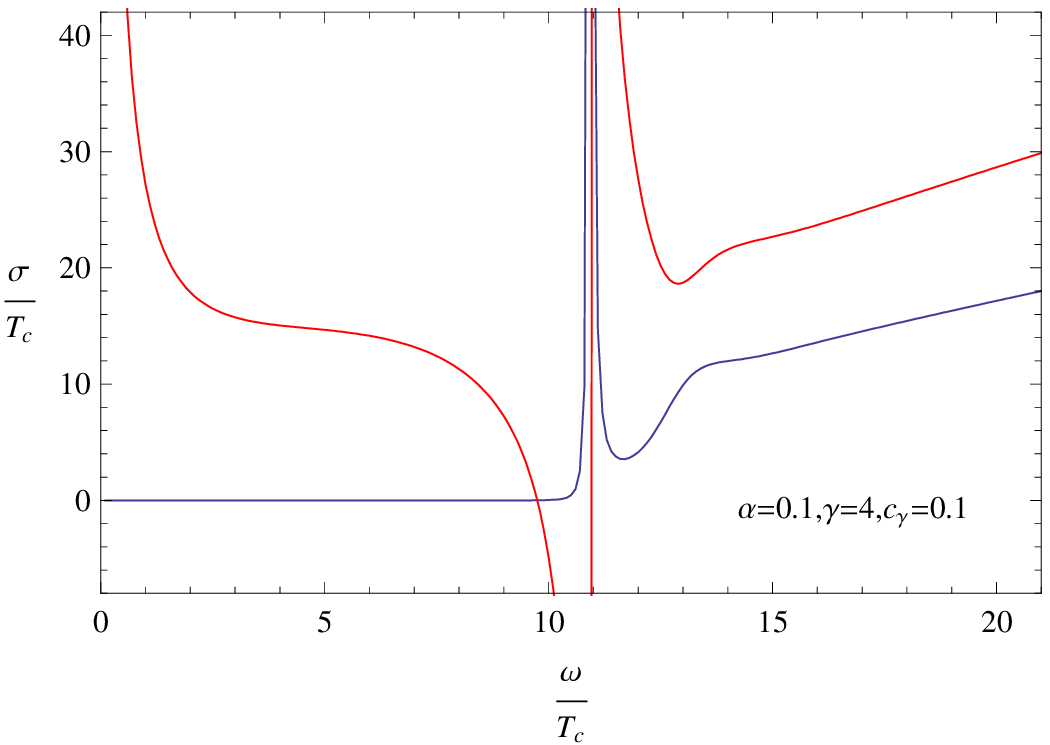}\\ \vspace{0.0cm}
\caption{\label{Conductivity} (color online) Conductivity for
($3+1$)-dimensional Gauss-Bonnet superconductors with fixed values
of $\alpha$ for different models with
$\mathfrak{F}(\psi)=\psi^{2}+c_{\gamma}\psi^{\gamma}$. The blue
(bottom) line and red (top) line represent the real part and
imaginary part of the conductivity respectively.}
\end{figure}

In \cite{HorowitzPRD78} it was argued that there is a universal
relation between the gap $\omega_g$ in the frequency dependent
conductivity and the critical temperature $T_c$: $\omega_g/T_{c}
\approx 8$ respected to a good approximation by all cases
considered. However this claimed universal relation was challenged
when the higher curvature corrections are taken into account
\cite{Gregory,Pan-Wang}. The available discussions were in the model
when $\mathfrak{F}(\psi)=\psi^{2}$. In the general $\mathfrak{F}$
form in our work, we will show that the model parameters $\gamma,
c_{\gamma}$ also modify the claimed universal relation even when
$\alpha=0$. In Fig. \ref{Conductivity} we plot the conductivity at
temperature $T/T_{c}\simeq0.3$. The blue (bottom) line represents
the real part of the conductivity and the red (top) line is the
imaginary part. We can easily find a gap in the conductivity with
the gap frequency $\omega_{g}$ changes with the values of
Gauss-Bonnet correction term $\alpha$, model parameters $\gamma$ and
$c_{\gamma}$. Fixing $\gamma$ and $c_{\gamma}$, the gap frequency
$\omega_g$ becomes larger for bigger $\alpha$, which agrees with the
finding in \cite{Gregory,Pan-Wang}. The gap frequency $\omega_g$
increases with $\gamma$ for fixed $c_{\gamma}, \alpha$ and grows
with $c_{\gamma}$ for selected $\gamma, \alpha$. The deviation from
$\omega_g/T_{c} =8$ becomes bigger with the increase of $\gamma$ and
$c_{\gamma}$. This holds even when $\alpha=0$. This shows that not
only the high curvature correction, but also the form of the scalar
field $\mathfrak{F}$ will affect the so-called universal relation
$\omega_g/T_{c}\approx8$ \cite{HorowitzPRD78}.

\section{conclusions}

We have introduced a general class of gravity dual with Gauss-Bonnet
corrections to describe both first and second order phase
transitions at finite temperature in strongly interacting systems.
In the probe limit, we found that besides the parameters which
define $\mathfrak{F}$ can separate the first and second order phase
transitions, different values of the Gauss-Bonnet correction bring
richer physics and can also change the order of the phase transition
in the generalized system. However for the second order phase
transition, we observed that the shift of the critical exponents
from that of the mean field result only appears for the parameters
defining $\mathfrak{F}$ and is independent of the Gauss-Bonnet
coupling. This is different from what we observed for the critical
temperature $T_c$, which only depends on the Gauss-Bonnet constant
while has nothing to do with other model parameters.

We also discussed the influences of the Gauss-Bonnet corrections and
other model parameters on the conductivity. We found that the size
and strength of the coherence peak can not only be controlled by the
parameters that define $\mathfrak{F}$ as observed in \cite{Franco},
but also be influenced by the Gauss-Bonnet coupling. This shows that
$\mathfrak{F}$ together with the high curvature correction controls
the magnitude of the fluctuations in the system. Furthermore we
examined the relation between the gap $\omega_g$ in the frequency
dependent conductivity and the critical temperature $T_c$. We found
that in addition to the high curvature influence on the ratio
$\omega_g/T_c$ observed in \cite{Gregory,Pan-Wang}, the model
parameters defining $\mathfrak{F}$ also give corrections to the
so-called universal relation $\omega_g/T_c\approx 8$.

\begin{acknowledgments}

This work was partially supported by the National Natural Science
Foundation of China. Qiyuan Pan was also supported by the China
Postdoctoral Science Foundation.

\end{acknowledgments}

\end{document}